\documentclass{elsart}

\usepackage{epsfig}
\usepackage{graphicx}
\usepackage{amssymb}
\usepackage{amsmath}
\begin{document}
\begin{frontmatter}

\title{Nonextensive Nuclear Liquid-Gas \\ Phase Transition}

\author{A. Lavagno, D. Pigato}
\address{Department of Applied Science and Technology, Politecnico di Torino, Italy}
\address{Istituto Nazionale di Fisica Nucleare (INFN), Sezione di Torino, Italy}

\maketitle

\begin {abstract}
We study an effective relativistic mean-field model of nuclear matter with arbitrary proton fraction at finite temperature in the framework of nonextensive statistical mechanics, characterized by power-law quantum distributions. We investigate the presence of thermodynamic instability in a warm and asymmetric nuclear medium and study the consequent nuclear liquid-gas phase transition by requiring the Gibbs conditions on the global conservation of baryon number and electric charge fraction. We show that nonextensive statistical effects play a crucial role in the equation of state and in the formation
of mixed phase also for small deviations from the standard Boltzmann-Gibbs statistics.

\vspace{0.5cm}
\end {abstract}

\end{frontmatter}

\section{Introduction}

The study of the thermodynamic properties of strongly interacting nuclear matter and the related phase transitions under extreme conditions is one of the most important goal of heavy ion collision experiments at intermediate and high energies. At low temperatures ($T\le 20$ MeV) and subnuclear densities, a liquid-gas type of phase transition was first predicted theoretically \cite{kapusta} and later observed experimentally in a nuclear multifragmentation phenomenon at intermediate-energy nuclear reactions \cite{pocho,xu}.

Because nuclei are made of neutrons and protons, the nuclear liquid-gas phase transition is in a binary
system where one has to deal with two independent proton and neutron chemical potentials for baryon number and electric charge  conservation. In fact, the information coming from experiments with heavy ions in intermediate- and high-energy collisions is that the Equation of State (EOS) depends on the energy beam but also sensibly on the proton fraction $Z/A$ (or isospin density) of the colliding nuclei \cite{ditoro2006}. Moreover, the study of nuclear matter with arbitrary electric charge fraction turn out to be important in radioactive beam experiments and in the physics of compact stars. Taking into account this important property, a very detailed study by M\"uller and Serot \cite{mullerserot} focused on the main thermodynamic properties of asymmetric nuclear matter in the framework of a relativistic mean field model.
Examples of other two-component systems are binary alloys and liquid $^3$He with spin-up and spin-down fluids.

A relevant aspect of a system with two conserved charges in asymmetric nuclear matter, is that, at variance with the so-called Maxwell construction for one conserved charge, the pressure is not constant in the mixed phase and therefore the incompressibility does not vanish \cite{mullerserot,prl2007}. Another interesting aspect of two-component systems is the possibility of having different proton-neutron ratios in the liquid and gas phases because of the symmetry energy, while still conserving the overall initial proton fraction (or isospin density).
Moreover, for a binary system with two phases, the binodal coexistence surface is two dimensional and the instabilities in the mixed liquid-gas phase arise from fluctuations in the baryon density (mechanical instability) and in the proton concentration (chemical instability) \cite{mullerserot,baran,prc2012}.
Such a feature plays also a crucial role in the structure and in the possible hadron-quark phase transition in compact star objects \cite{glendenning,astro}.

Recently, there has been increasing evidence that the nonextensive statistical mechanics, proposed by Tsallis, can be considered as an appropriate basis to deal with physi\-cal phenomena where strong dynamical correlations, long-range interactions and
microscopic memory effects take place \cite{tsallis,GMTsallis,book2,kodama,tsallis_new}. A considerable variety of
physical applications involve a quantitative agreement between experimental data and theoretical models based on Tsallis'
thermostatistics. In particular, in the last years there has been a growing interest in high energy physics applications of
nonextensive statistics and several authors have outlined the possibility that experimental observations in
relativistic heavy ion collisions can reflect nonextensive statistical behaviors
\cite{bediaga,albe,beck,wilk1,plb2001,biroprl05,physicaA2008,cley,chinellato}.

Nonextensive statistical mechanics introduced by Tsallis consists of a generalization of the common Boltzmann-Gibbs statistical mechanics and it is based upon the introduction of entropy \cite{book2,silva2}
\begin{eqnarray}
S_q[f]=\frac{1}{q-1}\, \left(1-\int[f({\bf x})]^q
\,d\Omega\right)\; ,\ \ \ \left(\int f({\bf x})\,d\Omega=1\right)
\, , \label{eq:GMTsallis}
\end{eqnarray}
where $f({\bf x})$ stands for a normalized probability distribution, ${\bf x}$ and $d\Omega$ denoting, respectively, a generic point and the volume element in the corresponding phase space (here and in the
following we set the Boltzmann and the Planck constant equal to unity). The real parameter $q$ determines the degree of non-additivity exhibited by the entropy form (\ref{eq:GMTsallis}) which reduces to the standard Boltzmann-Gibbs entropy in the limit $q\rightarrow 1$.
By means of maximizing the entropy $S_q$, under appropriate constraints, it is possible to obtain a probability distribution (or particle distribution) which generalized, in the classical limit, the Maxwell-Boltzmann distribution.
The nonextensive classical distribution can be seen as a superposition of the Boltzmann one with different temperatures which has a mean value corresponding to the temperature appearing in the Tsallis distribution \cite{wilk1,cley}.

The existence of nonextensive statistical effects strongly affects the finite temperature and density nuclear EOS \cite{physicaEOS,jpg,silva,wilknjl}. In fact, by varying temperature and density, the EOS reflects in terms of the macroscopic thermodynamical variables the microscopic interactions of the different phases of nuclear matter. The extraction of information about the EOS at different densities and temperatures by means of heavy ion collisions is a very difficult task and can be realized only indirectly by comparing the experimental data with different theoretical models, such as, for example, fluid-dynamical models. Related to this aspect, it is relevant to observe that a relativistic kinetic nonextensive theory \cite{pla2002} and a nonextensive version of a hydrodynamic model for multiparticle production processes have been proposed \cite{wilk08}.

In this paper we are going to study the influence of nonextensive statistical effects on the thermodynamical instabilities in warm and asymmetric nuclear matter and we investigate how the phase diagram of the nuclear liquid-gas phase transition can be modified in the framework of nonextensive statistical mechanics.

\section{Nonextensive hadronic equation of state}

The basic idea of the relativistic mean field model, widely and successful used for describing the properties of finite nuclei as well as hot and dense nuclear matter \cite{walecka,boguta,serot}, is the interaction between baryons through the exchange of mesons. In the original version, we have an isoscalar-scalar $\sigma$ meson field which produces the medium range attraction and the exchange of isoscalar-vector $\omega$ mesons responsible for the short range repulsion. The saturation density and binding energy per nucleon of nuclear matter can be fitted exactly in the simplest version of this model but other properties of nuclear matter, for example, incompressibility, cannot be well reproduced. To overcome these difficulties, the model has been modified introducing in the Lagrangian two terms of self-interaction for the $\sigma$ mesons that are crucial to reproduce the empirical incompressibility of nuclear matter and the effective mass of nucleons $M^*_N$. Moreover, the introduction of an isovector-vector $\rho$ meson allows to reproduce the correct value of the empirical symmetry energy.

In this framework, the Lagrangian density describing the nucleonic degrees of freedom ($p,n$) can be written as \cite{glen}
\begin{eqnarray}\label{eq:1}
{\mathcal L}_{N}&=&
\bar{\psi}[i\gamma_{\mu}\partial^{\mu}-(M- g_{\sigma}\sigma)
-g{_\omega}\gamma_\mu\omega^{\mu}-g_\rho\gamma^{\mu}\vec\tau\cdot
\vec{\rho}_{\mu}]\psi
+\frac{1}{2}(\partial_{\mu}\sigma\partial^{\mu}\sigma-m_{\sigma}^2\sigma^2)
\nonumber\\
&&-U(\sigma)+\frac{1}{2}m^2_{\omega}\omega_{\mu}\omega^{\mu}
+\frac{1}{2}m^2_{\rho}\vec{\rho}_{\mu}\cdot\vec{\rho}^{\;\mu}
-\frac{1}{4}F_{\mu\nu}F^{\mu\nu}
-\frac{1}{4}\vec{G}_{\mu\nu}\vec{G}^{\mu\nu}\,,
\end{eqnarray}
and $M=939$ MeV is the vacuum nucleon mass. The field strength
tensors for the vector mesons are given by the usual expressions
$F_{\mu\nu}\equiv\partial_{\mu}\omega_{\nu}-\partial_{\nu}\omega_{\mu}$,
$\vec{G}_{\mu\nu}\equiv\partial_{\mu}\vec{\rho}_{\nu}-\partial_{\nu}\vec{\rho}_{\mu}$,
and $U(\sigma)$ is a nonlinear potential of $\sigma$ meson
\begin{eqnarray}
U(\sigma)=\frac{1}{3}a\sigma^{3}+\frac{1}{4}b\sigma^{4}\,,
\end{eqnarray}
usually introduced to achieve a reasonable compression modulus for equilibrium nuclear matter.

The field equations in a mean field approximation are
\begin{eqnarray}
&&(i\gamma_{\mu}\partial^{\mu}-(M- g_{\sigma}\sigma)-
g_\omega\gamma^{0}\omega-g_\rho\gamma^{0}{\tau_3}\rho)\psi=0\,, \\
&&m_{\sigma}^2\sigma+ a{{\sigma}^2}+ b{{\sigma}^3}=
g_\sigma<\bar\psi\psi>=g_\sigma{\rho}_S\,, \\
&&m^2_{\omega}\omega=g_\omega<\bar\psi{\gamma^0}\psi>=g_\omega\rho_B\,,\\
&&m^2_{\rho}\rho=g_\rho<\bar\psi{\gamma^0}\tau_3\psi>=g_\rho\rho_I\,,
\label{eq:MFT}
\end{eqnarray}
where $\sigma=\langle\sigma\rangle$,
$\omega=\langle\omega^0\rangle$ and $\rho=\langle\rho^0_3\rangle$
are the nonvanishing expectation values of meson fields, $\rho_I=\rho_p-\rho_n$
is the isospin density, $\rho_B$ and $\rho_S$ are the baryon
density and the baryon scalar density, respectively. They are
given by
\begin{eqnarray}
&&\rho_{B}=2 \sum_{i=n,p} \int\frac{{\rm
d}^3k}{(2\pi)^3} \, n^q_i(k)\,, \label{eq:rhob} \\
&&\rho_S=2 \sum_{i=n,p} \int\frac{{\rm
d}^3k}{(2\pi)^3}\,\frac{M_i^*}{E_i^*} \, n_i^q(k)\,, \label{eq:rhos}
\end{eqnarray}
where $n_i(k)$ are the $q$-deformed fermion (proton and nucleon) particle distributions (the antiparticle degrees of freedom are negligible in the range of temperature and density explored in this paper). More explicitly, for $q>1$ and $\beta(E_i^*-\vert\mu_i^*\vert)>0$, we have \cite{cley,wilknjl}
\begin{eqnarray}
n_i(k)=\frac{1} { [1+(q-1)\,\beta(E_i^*(k)-\mu_i^*)
]^{1/(q-1)} + 1} \label{eq:distribuz} \, .
\end{eqnarray}

The nucleon effective energy is defined as
${E_i}^*(k)=\sqrt{k^2+{{M_i}^*}^2}$, where ${M_i}^*=M_{i}-g_\sigma
\sigma$. The effective chemical potentials $\mu_i^*$  are given in
terms of the meson fields as follows
\begin{eqnarray}
\mu_i^*={\mu_i}-g_\omega\omega -\tau_{3i} g_{\rho}\rho \, ,
\label{mueff}
\end{eqnarray}
where $\mu_i$ are the thermodynamical chemical potentials
$\mu_i=\partial\epsilon/\partial\rho_i$. At zero temperature, they
reduce to the Fermi energies $E_{Fi} \equiv
\sqrt{k_{Fi}^2+{M_i^*}^2}$ and the nonextensive statistical effects
disappear.

Because we are going to describe a finite temperature and density asymmetric nuclear matter, we have to require the conservation of two "charges": baryon number ($B$) and electric charge ($C$) (we neglect the contribution of strange hadrons, because a very tiny amount of strangeness can be produced in the range of temperature and density explored in this study) \cite{prc2010}. As a consequence, the system is described by two independent chemical potentials: $\mu_B$ and $\mu_C$, the baryon and the electric charge chemical potential, respectively. Therefore, the chemical potential of nucleon of index $i$ ($i=p,n$) can be written as
\begin{equation}
\mu_i=b_i\, \mu_B+c_i\,\mu_C \, , \label{mu}
\end{equation}
where $b_i$ and $c_i$ are, respectively, the baryon and the
electric charge quantum numbers of the nucleon under consideration.

The meson fields are obtained as a solution of the field
equations in mean field approximation and the related meson-nucleon
couplings ($g_\sigma$, $g_\omega$ and $g_\rho$) are the free
parameters of the model. In the following, they will be fixed to the
parameters set marked as TM1 of Ref.\cite{toki}.

The thermodynamical quantities can be obtained from the
thermodynamic potential in the standard way. More explicitly,
the baryon pressure $P_B$ and the energy density $\epsilon_B$ can
be written as
\begin{eqnarray}
&&P_B=\frac{2}{3} \sum_{i=n,p} \int \frac{{\rm
d}^3k}{(2\pi)^3} \frac{k^2}{E_{i}^*(k)}
[n_i^q(k)+\overline{n}_i^q(k)] -\frac{1}{2}m_\sigma^2\sigma^2 -
U(\sigma)\nonumber\\
&&\hspace{1cm}+\frac{1}{2}m_\omega^2\omega^2+\frac{1}{2}m_{\rho}^2 \rho^2\,,\label{eq:eos}\\
&&\epsilon_B= 2 \sum_{i=n,p}\int \frac{{\rm
d}^3k}{(2\pi)^3}E_{i}^*(k) [n_i^q(k)+\overline{n}_i^q(k)]
+\frac{1}{2}m_\sigma^2\sigma^2+U(\sigma) \nonumber\\
&&\hspace{1cm}+\frac{1}{2}m_\omega^2\omega^2+\frac{1}{2}m_{\rho}^2 \rho^2\, .
\label{eq:eos2}
\end{eqnarray}

\section{Nonextensive statistical effects in the nuclear liquid-gas phase transition}

We are dealing with the study of a multi-component system at finite temperature and density with two conserved charges: baryon number and electric charge. For such a system, the Helmholtz free energy density $F$ can be written as
\begin{equation}
F(T,\rho_B,\rho_C)= -P(T,\mu_B,\mu_C) +\mu_B\rho_B + \mu_C\rho_C \, ,
\end{equation}
with
\begin{equation}
\mu_B=\left (\frac{\partial F}{\partial\rho_B}\right)_{T,\,\rho_C} \, , \ \ \ \mu_C=\left ( \frac{\partial F}{\partial\rho_C}\right)_{T,\,\rho_B} \, .
\end{equation}

In general, a system can exist in a number of different phases, each of which exhibit quite different macroscopic behavior. The single phase that is realized for a given set of independent variables is the one with the lowest free energy. In a system with two conserved charges, it is possible to have $N_{\rm max}=4$ phase coexistence regions in thermodynamical equilibrium  \cite{landau,reichl}, even if we have found no evidence for the existence of more than two phases in the regime investigated in this paper. Therefore,
assuming the presence of two phases (denoted as $I$ and $II$, respectively), the system is stable against the separation in two phases if the free energy of a single phase is lower than the free energy in all two phases configuration. The phase coexistence is given by the Gibbs conditions
\begin{eqnarray}
&&\mu_B^{I} = \mu_B^{II} \, , \ \ \ \ \ \ \ \ \ \mu_C^{I} = \mu_C^{II}
\, , \\
&&P^I (T,\mu_B,\mu_C)=P^{II} (T,\mu_B,\mu_C) \, .
\end{eqnarray}
At a given baryon density $\rho_B$ and a given net
electric charge density $\rho_C=y\, \rho_B$ (with $y=Z/A$), the chemical potentials $\mu_B$ are $\mu_C$ are univocally determined by the
following equations
\begin{eqnarray}
&&\!\!\!\!\!\!\!\!\rho_B=(1-\chi)\,\rho_B^I(T,\mu_B,\mu_C) +\chi \,\rho_B^{II}(T,\mu_B,\mu_C) \, ,\label{rhobchi}\\
&&\!\!\!\!\!\!\!\!\rho_C=(1-\chi)\,\rho_C^I(T,\mu_B,\mu_C) +\chi \,\rho_C^{II}(T,\mu_B,\mu_C) \, ,
\label{rhocchi}\end{eqnarray}
where $\rho_B^{I(II)}$ and $\rho_C^{I(II)}$ are, respectively, the baryon and electric charge densities in the low density ($I$) and in the higher density ($II$) phase and $\chi$ is the volume fraction of the phase $II$ in the mixed phase ($0\le\chi\le 1$).

An important feature of these conditions is that, unlike the case of a single conserved charge, the pressure in the mixed phase is not constant and, although the total $\rho_B$ and $\rho_C$ are fixed, baryon and charge densities can differ in the two phases, according to Eq.s (\ref{rhobchi}) and (\ref{rhocchi}).

For such a system in thermal equilibrium, the possible phase transition can be characterized by mechanical (fluctuations in the baryon density) and chemical instabilities (fluctuations in the electric charge density).
As usual, the condition of the mechanical stability implies
\begin{eqnarray}
\rho_B \left(\frac{\partial P}{\partial \rho_B}\right)_{T,\,\rho_C} >0   \, .  \label{InstabMecc}
\end{eqnarray}
By introducing the notation $\mu_{i,j}=(\partial\mu_i/\partial\rho_j)_{T,P}$ (with $i,j=B,C$), the chemical stability can be expressed with the following conditions \cite{reichl}
\begin{eqnarray}
\mu_{B,B} >0 \, , \ \ \ \mu_{C,C}>0 \, , \ \ \
\begin{vmatrix}
\,\mu_{B,B} & \mu_{B,C} \\
\,\mu_{C,B} & \mu_{C,C}
\end{vmatrix}
>0  \,  . \label{InstabChim}
\end{eqnarray}
In addition to the above conditions, for a process at constant $P$ and $T$, it is always satisfied that
\begin{eqnarray}
&&\rho_B\, \mu_{B,B}+\rho_C \, \mu_{C,B}=0\, , \\
&&\rho_B\, \mu_{B,C}+\rho_C \, \mu_{C,C}=0\, . \label{diff1}
\end{eqnarray}

Whenever the above stability conditions are not respected, the system becomes unstable and the phase transition take place.  The coexistence line of a system with one conserved charge becomes in this case a two dimensional surface in $(T,P,y)$ space, enclosing the region where mechanical and chemical instabilities occur.

In a regime of low temperature and baryon density, relevant in the liquid-gas phase transition, only proton and neutron degrees of freedom take place and $\mu_B=\mu_n$, $\mu_C=\mu_p-\mu_n$, $\rho_B=\rho_p+\rho_n$,  $\rho_C=\rho_p$.
In this simple case, for example, Eq. (\ref{diff1}) can be written as
\begin{equation}
y\, \left(\frac{\partial \mu_p}{\partial y}\right)_{T,P}+(1-y)\, \left(\frac{\partial \mu_n}{\partial y}\right)_{T,P}=0 \, ,\label{diff2}
\end{equation}
where $y=\rho_p/\rho_B$. Because we are working with a proton fraction $0< y\le 0.5$, the chemical stability conditions (\ref{InstabChim}) are therefore satisfied if
\begin{equation}
\left(\frac{\partial \mu_p}{\partial y}\right)_{T,P}>0  \ \ {\rm or } \ \ \left(\frac{\partial \mu_n}{\partial y}\right)_{T,P}<0 \, .
\label{inst_lg}
\end{equation}
The chemical instability region boundaries are determined by the points where the slope of each chemical potential with respect to $y$ is zero.

We are now able to study the effects of nonextensive statistical effects on the presence of nuclear instabilities at low temperature and subnuclear densities, in the framework of the previous introduced nuclear EOS. In the following we will focus our investigation for small deviations from the standard Boltzmann-Gibbs statistical mechanics and for values $q>1$, because these values were obtained in several phenomenological studies in high energy heavy ion collisions (see, for example, Ref.s \cite{physicaA2008,cley,chinellato,physicaEOS}).  In this context, let us remember that, in the diffusional approximation, a value $q>1$ implies the presence of a superdiffusion among the constituent particles (the mean square displacement obeys to a power law behavior $\langle x^2\rangle\propto t^\alpha$, with $\alpha>1$) \cite{tsamem}.

\begin{figure}
\begin{center}
\resizebox{1.\textwidth}{!}{%
\includegraphics{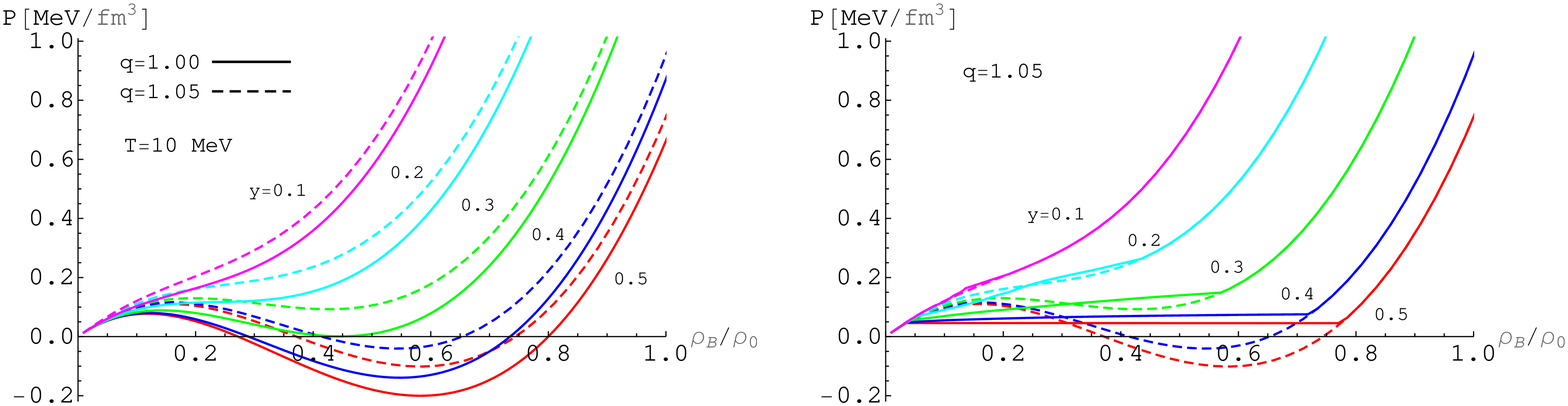}
} \caption{Pressure as a function of baryon density (in units of the nuclear saturation density $\rho_0$) for various values of the proton fraction $y$ and at $T=10$ MeV. In the left panel, the dashed lines are related to the entropic value $q=1.05$ in comparison to the standard case ($q=1$), corresponding to the continuous line. In the right panel, the continuous (dashed) lines correspond to the solution obtained with (without) the Gibbs construction for $q=1.05$.} \label{fig:P_rhob_lg}
\end{center}
\end{figure}

In Fig. \ref{fig:P_rhob_lg}, we report the pressure as a function of baryon density $\rho_B$ (in units of the nuclear saturation density $\rho_0$) for various values of the electric charge fraction $y$ at fixed temperature $T=10$ MeV.
In the left panel, a comparison of the equation of state in the presence of ($q=1.05$, dashed lines) and in the absence ($q=1$, continuous lines) of the nonextensive statistical effects is shown. In the right panel, for $q=1.05$ only, the continuous lines correspond to the solution obtained with the Gibbs construction in the mixed phase, whereas the dashed lines are without correction.

Let us observe that in presence of small deviation from the standard Boltzmann-Gibbs statistics, the nuclear EOS appears stiffer, with higher values of pressure at fixed baryon density. As we will see, this implies a greater value of the first transition density and a reduction of second transition density in the mixed phase with respect to the standard case ($q=1$). This feature results in significant changes in the nuclear incompressibility and may be particularly important in identifying the presence of nonextensive effects in heavy-ion collision experiments. Moreover, by remembering Eq.(\ref{InstabMecc}), it appears evident that for a proton fraction $y>0.2$ a mechanical instability is present, whereas for $y<0.2$ the system becomes unstable only under chemical-diffusive instability.

\begin{figure}
\begin{center}
\resizebox{0.7\textwidth}{!}{%
\includegraphics{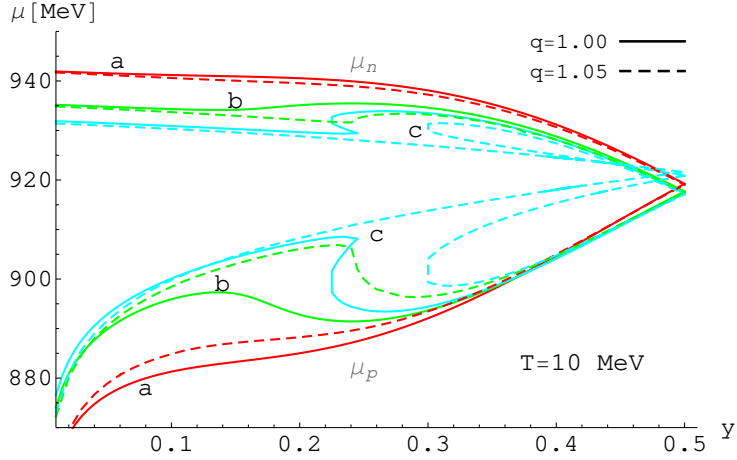}
} \caption{Proton and neutron chemical potential as a function of the proton fraction $y$ for various isobars ($P$=0.35, 0.15, 0.10 MeV/fm$^3$) (lines $a$ to $c$) at $T$=10 MeV and for different values of $q$.
} \label{fig:mu-za}
\end{center}
\end{figure}

The chemical unstable regions, in the presence and in absence of nonextensive statistical effects, are much more evident in Fig. \ref{fig:mu-za}, where we show the proton and neutron chemical potentials for various isobars at fixed temperature $T=10$ MeV, as a function of the proton asymmetry.
Below $P=0.35$ MeV/fm$^3$, the system becomes unstable because of the presence of regions of negative (positive) slope for $\mu_p$ ($\mu_n$) and the relevance of chemical instability becomes more important in presence of nonextensive statistical effects. By increasing the value of $q$, the chemical instabilities become much more pronounced and the system tends to become numerically unstable.

\begin{figure}
\begin{center}
\resizebox{0.7\textwidth}{!}{%
\includegraphics{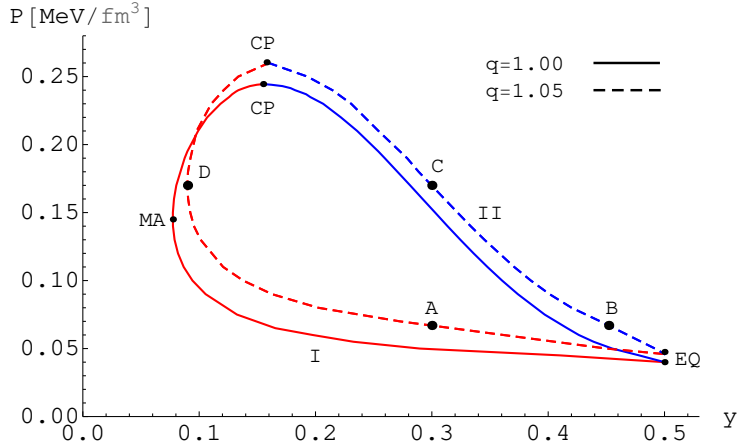}
} \caption{Binodal section at $T=10$ MeV and for different values of $q$, with evidence of the critical point (CP), the point of maximum asymmetry (MA) and the point of equal equilibrium (EQ).} \label{fig:P_ZA}
\end{center}
\end{figure}

In order to study the phase coexistence of the system, in Fig. \ref{fig:P_ZA}, we show the binodal section as a function of the proton asymmetry $y$ at $T=10$ MeV, in the presence ($q=1.05$) and in the absence ($q=1$) of nonextensive statistical effects.
Following the same notation as Ref. \cite{mullerserot}, the binodal surface is divided into two branches by a critical point (CP), beyond which ends the phase transition, and a point of equal equilibrium (EQ) at $y=0.5$, where protons and neutrons have the same concentration. The left branch of the diagram represents the initial phase configuration of the system at lower density (gas phase, I) and the second branch, at higher density, corresponds to the final phase configuration (liquid phase, II). The binodal surface encloses the area where the system undergoes the phase transition.

During the isothermal compression, the system evolves through configuration at constant proton fraction $y_A$ and meets the first branch in a generic point $A$. At this point the system becomes unstable and an infinitesimal phase in $B$ appears at the same temperature and pressure of $A$. In this context let us remember that, although the proton asymmetry is globally conserved, this is not true for the single phase. In particular, for an asymmetric nuclear system ($y\ne 0.5$) it is energetically favorable to separate it into a liquid phase (less asymmetric, $y>y_A$) and a gas phase (more asymmetric, $y<y_A$) rather than into two phases with equal proton fraction. As the system is compressed, during the phase transition ($0 <\chi< 1$), the total proton fraction $y$ remains fixed but each phase evolves towards a configuration with different $y$. The gas phase evolves from $A$ to $D$, while the liquid phase evolves from $B$ to $C$. At the point $C$, the system leaves the region of instability.

If point $A$ has a value of $y_A$ greater than the corresponding values $y_{\rm CP}$ of the CP, the system ends the phase transition in the liquid phase (in the point $C$).
On the other hand, as already observed in Ref. \cite{mullerserot}, if the system has been prepared in a very asymmetric configuration with $y_{A^{'}}< y_{\rm CP}$, it undergoes to a retrograde phase transition. A second liquid phase in $B^{'}$ is formed but, after reaching a point of maximum volume fraction $\chi_{\rm max}<1$, the system returns to its initial gas phase at point $C^{'}$. Note that this kind of phase transition is possible only for a multi-component system and, in this case, corresponds to a purely chemical-diffusive instability.

Let us note that in the presence of a nonextensive statistical effect, for $y_A\ge 0.3$ (in the range of more physical interest because, below such a value it is very difficult to test nuclear matter, also with radioactive ion beam facilities), the proton fractions $y_i$ of the two phases have values closer to the initial value $y_A$. In other words, the phase $I$ turns out to be less asymmetric (greater values of $y$) and the phase $II$ less symmetric (lower values of $y$) with respect to the standard case. In the liquid-gas phase transition, the process of producing a larger neutron excess in the gas phase (low values of $y$) is referred to as isospin fractionation or neutron distillation. This effect results to be lessened in presence of nonextensive statistical effects. In this context, it is important to remember that a similar feature is present when Coulomb interaction is included the nuclear medium, which leads to proton diffusion of some protons from the liquid phase back into the gas phase \cite{baran}.

\begin{figure}
\begin{center}
\resizebox{0.7\textwidth}{!}{%
\includegraphics{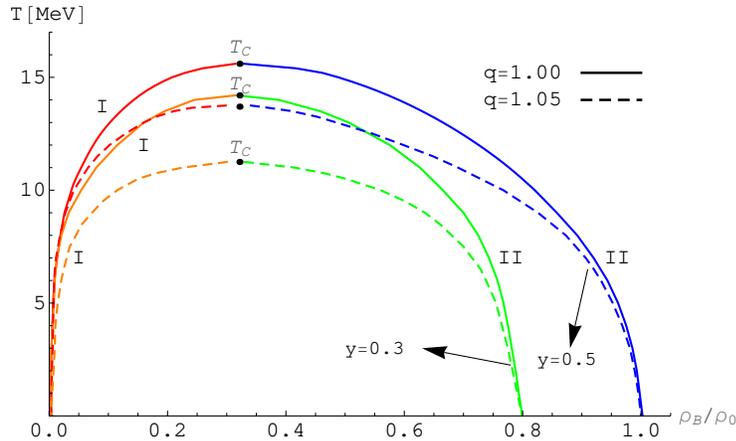}
} \caption{Phase diagram of the liquid-gas phase transition for asymmetric ($y=0.3$) and symmetric ($y=0.5$) nuclear matter and for different values of $q$. The lines labeled with I and II, delimitate the first and second critical densities of the coexistence regions, respectively.} \label{fig:T_rhobtot}
\end{center}
\end{figure}

Finally, in Fig. \ref{fig:T_rhobtot}, we report the phase diagram with evidence of the coexi\-stence regions of the liquid-gas phase transition for $y=0.3$ and $0.5$, in the pre\-sence ($q=1.05$) and in the absence ($q=1$) of nonextensive statistical effects. As expected, the relevance of nonextensive statistical mechanics increases by increasing the temperature and implies a reduction of  the critical temperature $T_c$ and a reduction of the baryon density range involved in the mixed phase, with an enhancement of the first critical density and a reduction of the second critical density, with respect to the standard (extensive) case.

\section{Conclusions}
We have studied an effective nuclear EOS in the framework of nonextensive statistical effects at finite temperature and baryon density,  by requiring the Gibbs conditions on the global conservation of the baryon number and electric charge (proton) fraction.  The study of a two-component nuclear system with an arbitrary proton/neutron ratio turns out to be very important for RIB (Rare Isotope Beam) Facility experiments and in astrophysical investigation such as in neutron stars. A novel aspect of a binary asymmetric nuclear system is the possibility of having different proton-neutron ratios in each of the phases, while still conserving the overall initial proton fraction. Due to this higher dimensionality, the phase diagram is now a surface in the pressure, temperature and in the proton fraction $y$ or baryon density.

We have shown that several features of the nuclear liquid-gas phase transition turn out to be sensibly modified also for small deviations from the standard Boltzmann-Gibbs statistics. In asymmetric nuclear matter, instabilities that produce a liquid-gas phase separation arise from fluctuations in the proton fraction (chemical instability), rather than from fluctuations in the baryon density (mechanical instability). In the presence of nonextensive statistical effects, the relevance of the chemical instability becomes more pronounced; it appears an effective reduction of the dynamical instability region and the so-called
isospin fractionation effect turns out to be reduced, with fewer protons in the liquid phase and more proton in the gas phase, with respect to the standard case. Such features have been discussed in literature when a more realistic nuclear system is studied with the inclusion of long range forces such as Coulomb interactions. In this sense, nonextensive statistical effects can phenomenologically take into account the complex many-body interaction in the nuclear medium where dynamical strong correlations and long-range forces are present.

\end{document}